\documentclass[12pt]{article}
 
\def\be{\begin{equation}} 
\def\ee{\end{equation}} 

\def\ch{\mbox{ch}}
\def\sh{\mbox{sh}}
\def\sign{\mbox{sign}}
\def\tg{\mbox{tg}}         
\def\s{\sigma} 
\def\g{\gamma}
\def\a{\alpha}
\def\b{\beta} 
\def\l{\lambda}
\def\la{\langle}
\def\ra{\rangle}

\begin{document}

\begin{center}
{\large Fisher-Hartwig conjecture and the correlators in 
XY spin chain.}
\end{center}

\begin{center}
{\large A.A.Ovchinnikov}
\end{center}

\begin{center}
{\it Institute for Nuclear Research, RAS, Moscow, 117312, Russia}  
\end{center}

\begin{abstract}

We apply the theorems from the theory of Toeplitz determinants 
to calculate the asymptotics of the correlators in the XY spin chain 
in the transverse magnetic field. 
The asymptotics of the correlators for the XX spin chain 
in the magnetic field are obtained.  

\end{abstract}

\vspace{0.2in}

{\bf 1. Introduction} 
\vspace{0.1in}

  The calculation of correlators in the one-dimensional exactly 
solvable models remains an interesting and important problem. 
An important exact results on the correlators for the XY 
quantum spin chain and the Ising model were obtained 
\cite{LSM}-\cite{MC}. 
    The goal of the present letter is twofold. 
  First, we present the calculations of the correlators in 
the limit of the XX spin chain in a non-zero magnetic field 
which was not considered previously. 
In particular, the asymptotics of the exponential correlator 
for the XX spin chain is found.  
  Second, we use the generalized Fisher-Hartwig conjecture 
(for example, see \cite{BM}) and the method to calculate 
the asymptotics of Toeplitz determinants \cite{F} to 
derive the known results \cite{MC} for the correlators 
in the XY spin chain in a non-zero magnetic field 
in a simple way.

In the present letter to calculate the asymptotics of the 
determinants we use the conjecture from the theory of Toeplitz 
matrices, \cite{BM}, the generalized Fisher-Hartwig conjecture,   
which is based on the proofs of the original 
Fisher-Hartwig conjecture \cite{FH} including the constant in 
front of the asymptotics, in a number of the particular cases 
\cite{Basor}, \cite{ES} (see also \cite{BM} and references 
therein). 
    One finds the representation of the generating function 
$f(x)$ of the Toeplitz matrix 
$M_{ij}=M(i-j)=\int_{0}^{2\pi}(dx/2\pi)e^{i(i-j)x}f(x)$ 
of the following form: 
\be
f(x)=f_0(x)\prod_{r}e^{ib_r(x-x_r-\pi\sign(x-x_r))}
(2-2\cos(x-x_r))^{a_r},  
\label{f}
\ee
where $x\in(0,2\pi)$ is implied, the discontinuities (jumps and zeroes 
or the power-law singularities) are at finite number of the points $x_r$,  
and $f_0(x)$ is the smooth non-vanishing function 
with the continuously defined argument at the interval $(0,2\pi)$.  
The function (\ref{f}) is 
characterized by the parameters $a_r$, $b_r$ at each point of  
discontinuity $x_r$. 
In general there are several representations of the form (\ref{f}) 
for a given functions $f(x)$. 
To obtain the asymptotics of the determinant one should take the sum 
over the representations (\ref{f}) corresponding to the minimal exponent 
$\sum_r(b_r^2-a_r^2)$: 
\be
D(N)=\sum_{Repr.}e^{l_0 N}N^{\sum_r(a_r^2-b_r^2)} E
\label{DN}
\ee
where $E$ is the constant independent of $N$, 
\be
E=\exp\left( \sum_{k=1}^{\infty}k l_k l_{-k}\right) 
\prod_{r}(f_{+}(x_r))^{-a_r+b_r}(f_{-}(x_r))^{-a_r-b_r}~~~~~~~~~~~~
\label{E}
\ee
\[
~~~~~~~~~\prod_{r\neq s}\left(1-e^{i(x_s-x_r)}\right)^{-(a_r+b_r)(a_s-b_s)}  
\prod_{r}\frac{G(1+a_r+b_r)G(1+a_r-b_r)}{G(1+2a_r)},  
\]
$l_k=\int_{-\pi}^{\pi}(dx/2\pi)e^{ikx}\ln(f_0(x))$, and 
the functions $f_{\pm}(x)$ are given by the equations  
\[
\ln f_{+}(x)=\sum_{k>0}l_{-k}e^{ikx}, ~~~
\ln f_{-}(x)=\sum_{k>0}l_{k}e^{-ikx},   
\]
where $G$ is the Barnes $G$-function \cite{Barnes}, 
$G(z+1)=G(z)\Gamma(z)$, $G(1)=1$.

For the piecewise continuous generation function $f(x)$ 
with the finite number of discontinuities which is not equal 
to zero ($a_r=0$) the equation (\ref{DN}) can be represented in 
a more convenient form \cite{Basor}. 
Suppose the function $f(x)$ has 
the finite number of discontinuities at the points $x_r$, 
\[
\l_r=\frac{1}{2\pi}\left(\ln f(x_r+0)-\ln f(x_r-0)\right), 
\]
and assume that the function $f(x)$ has the continuously defined 
argument and not equal to zero at the interval $(-\pi,\pi)$. 
Then the asymptotic of the determinant is given by the formula 
\cite{Basor} 
\be
D(N)=\sum_{Repr.}e^{l_0 N}N^{(\sum_r\l_r^{2})}E,~~~~E=\exp\left(  
\sum_{k=1}^{\infty}\left(k l_k l_{-k}-\frac{1}{k} 
\sum_{r}\l_r^{2}\right)\right)\prod_r \tilde{g}(\l_r),
\label{det}
\ee
where we denote 
$l_k=\int_{-\pi}^{\pi}(dx/2\pi)e^{ikx}\ln(f(x))$ 
and the function $\tilde{g}(\l)$ equals 
\be
\tilde{g}(\l)=e^{(1+\g)\l^2}\prod_{k=1}^{\infty}
\left(1+\frac{\l^2}{k^2}\right)e^{-\l^2/k}, 
\label{gl}
\ee
where $\g=0.577..$ is Euler's constant. 
The constant $E$ (\ref{gl}) in eq.(\ref{det}) is a consequence of 
the formula for the Barnes $G$- function: 
\[
G(1+z)=(2\pi)^{z/2}\exp\left(-\frac{z(z+1)}{2}-\g\frac{z^2}{2}\right) 
\prod_{k=1}^{\infty}\left(1+\frac{z}{k}\right)^k 
\exp(-z+z^2/2k), 
\]
in such a way that $\tilde{g}(\l_r)=G(1+z)G(1-z)$ for $\l_r=iz$.

In ref.\cite{Basor} the theorem was proved for an arbitrary 
number of the discontinuities of the imaginary part of the magnitude 
less than $1/2$, $|\l_r|<1/2$, however, there are many reasons 
to believe it to be true also in the case $|\l_r|=1/2$ \cite{BM}.  
For the case of an arbitrary single Fisher-Hartwig singularity 
the conjecture (\ref{DN}) was recently proved in ref.\cite{ES}.  
See ref.\cite{BM} for the complete list of the cases for which 
the rigorous proof of the conjecture (\ref{DN}) is available. 
      Schematically the rigorous proofs in the particular 
cases \cite{Basor} \cite{ES} go as follows.  
Suppose that equation (\ref{det}) is fulfilled for some functions 
$f_1(x)$ and $f_2(x)$ of the class of piecewise continuous 
functions with continuously defined argument. 
Then the equation (\ref{det}) is fulfilled for the function 
$f(x)=f_1(x)f_2(x)$. 
Thus it is sufficient to prove eq.(\ref{det}) for the 
smooth function with the continuously defined argument, 
in which case it is reduces to the well known 
strong Szego theorem and for the singular function of the 
form (\ref{f}) $f(x)=(1-z)^{\a}(1-1/z)^{\b}$, 
$z=e^{ix}$, in which case the asymptotics is known exactly.

\vspace{0.2in}

{\bf 2. General form of the correlators for XY spin chain.}

\vspace{0.1in}

Let us review the known calculations \cite{LSM}-\cite{MC} 
of the equal-time correlators of the operators $S^{\pm}$ 
for the XY spin chain at non-zero transverse magnetic field. 
The Hamiltonian of the XY spin chain has the form 
\be
H=-\sum_{i=1}^{L}\left((1+\g)S_i^{x}S_{i+1}^{x}+ 
(1-\g)S_i^{y}S_{i+1}^{y} - h S^{z}_{i}\right),
\label{H}
\ee
where $S^a=\frac{1}{2}\s^a$, $a=x,y,z$, $0<\g<1$   
and the periodic boundary conditions are implied. 
Performing the well-known Jordan-Wigner transformation
($S^{\pm}=S^{x}\pm iS^{y}$): 
\be
S_{x}^{+}=e^{i\pi N(x)}a_x^{+}=\exp(\sum_{l<x}n_{l})a_x^{+}, 
\label{J}
\ee
we obtain the following Hamiltonian written down in terms 
of the fermionic operators: 
\be 
H=-\frac{1}{2}\sum_{i=1}^{L}\left(a_i^{+}a_{i+1}+
\g a_i^{+}a_{i+1}^{+} +h.c.\right)+ 
h\sum_{i=1}^{L}a_i^{+}a_i, 
\label{HF}
\ee
where the boundary terms which are not important in the 
thermodynamic limit are omitted. 
The Hamiltonian is diagonalized with the help of the 
following Fourier transform in the sectors with an odd number 
of particles: 
\[
a_k^{+}=\frac{1}{\sqrt{L}}\sum_{x}e^{ikx}a_x^{+}, ~~~
k=\frac{2\pi n}{L}, ~~n\in Z, ~~~ -\pi<k<\pi. 
\]
The Hamiltonian (\ref{HF}) is easily diagonalized in the 
momentum space. Introducing the column 
$\psi_k^{+}=(a_k^{+}, a_{-k})$ for $k\in(0,\pi)$ the Hamiltonian 
is represented as 
\[
H=-\sum_{k>0}\psi_k^{+}\hat{M}_{k}\psi_k, ~~~~
\hat{M}_k=\left(
\begin{array}{cc}
\epsilon_k & i\Delta_k \\
-i\Delta_k & -\epsilon_k 
\end{array} \right), ~~~~\psi_k=\left( 
\begin{array}{c}
a_k \\
a^{+}_{-k}
\end{array}\right), ~~~~k>0, 
\]
where the notations $\epsilon_k=\cos(k)-h$, $\Delta_k=\g\sin(k)$ 
are used and the additional constant term $\epsilon_k$ is implied. 
It is convenient to redefine the operator $a_{-k}^{+}\to ia_{-k}^{+}$. 
In this case the Hamiltonian is diagonalized by means of the 
canonical transformation 
\be
a_{k}^{+}=c_{k}\a_{1k}^{+}+s_{k}\a_{2k}, ~~~
a_{-k}= c_{k}\a_{2k}-s_{k}\a_{1k}^{+}, 
\label{can}
\ee
where $k>0$, $c_k^2+s_k^2=1$, $c_k=\cos(\phi_k)$, $s_k=\sin(\phi_k)$, 
and the angle $\phi_k$ and the Hamiltonian are: 
\[
\tg(2\phi_k)=-\frac{\Delta_k}{\epsilon_k}, ~~~
H=-\sum_{k>0}\left(\a_{1k}^{+}\a_{1k}-\a_{2k}^{+}\a_{2k}+1\right), ~~~
E_k=\sqrt{(\cos(k)-h)^2+\g^2\sin^2(k)}. 
\]
Consider the equal-time correlators for the XY spin chain: 
\[
G(x)=\la0|S^{+}_{i+x}S^{-}_{i}|0\ra. ~~~~~
\tilde{G}(x)=\la0|S^{+}_{i+x}S^{+}_{i}|0\ra.
\]
Using the Jordan-Wigner transformation 
$S_{x}^{+}=\exp(i\pi\sum_{l<x}n_l)a_x^{+}$,  
the correlation functions $G(x)$, $\tilde{G}(x)$ can be represented 
as the following averages over the ground state of the Hamiltonian 
(\ref{HF}), 
\be
G(x)=\la0|a_x^{+}e^{i\pi N(x)}a_0|0\ra, ~~~~~
\tilde{G}(x)=-\la0|a_x^{+}e^{i\pi N(x)}a_{0}^{+}|0\ra, 
\label{corr}
\ee 
where $\hat{N}(x)=\sum_{i=1}^{x-1}n_{i}$. 
Introducing the operators, anticommuting at different sites, 
\[
A_i=a_i^{+}+a_i, ~~~~ B_i=a_i^{+}-a_i, ~~~~A_i B_i=e^{i\pi n_i},   
\]
where $n_{i}=a_i^{+}a_i$ - is the fermion occupation number, with the 
following correlators with respect to the vacuum, 
\[
\la B_i A_j\ra=BA(i-j),~~~~\la A_i A_j \ra=0,~~~\la B_i B_j\ra=0
\]
one obtains the following expression for the bosonic correlator: 
\be 
G(x)=\frac{1}{4}\la0|B_x\prod_{i=1}^{x-1}(A_i B_i)A_0|0\ra+ 
\frac{1}{4}\la0|B_0\prod_{i=1}^{x-1}(A_i B_i)A_0|0\ra = 
\la S^{y}_x S^{y}_0\ra + \la S^{x}_x S^{x}_0\ra, 
\label{G}
\ee
where the terms $\la S^{y}_x S^{y}_0\ra$, $\la S^{x}_x S^{x}_0\ra$
correspond respectively to the two terms at the 
left-hand side of this equation. At zero magnetic field 
each term in this expression is represented as a product of the two 
Toeplitz determinants \cite{LSM}. The asymptotic behaviour of $G(x)$ is 
determined by the form of the average $BA(x)$. To calculate this  
function one should use the expressions for the averages of two 
operators $a_i^{+}$, $a_i$, which follow from the canonical 
transformation (\ref{can}): 
\be
\la a_{x}^{+}a_0 \ra= \int_{0}^{\pi}\frac{dk}{2\pi}
\left( e^{ikx}+e^{-ikx}\right)\left(\frac{1}{2}+\frac{1}{2}
\frac{\cos(k)-h}{\sqrt{(\cos(k)-h)^2+\g^2\sin^2(k)}}\right), 
\label{av}
\ee
\[
\la a_x^{+} a_0^{+}\ra=\int_{0}^{\pi}\frac{dk}{2\pi}\left( 
e^{ikx}-e^{-ikx}\right)
\left(\frac{1}{2}\frac{i\g\sin(k)}{\sqrt{(\cos(k)-h)^2+\g^2\sin^2(k)}}     
\right)
\]
where the sums have been replaced by the integrals in the 
continuum limit. From the expressions (\ref{av}) 
the following expression can be obtained:  
\[
BA(x)=\int_{-\pi}^{\pi}\frac{dk}{2\pi}e^{ikx}
\frac{\cos(k)-h+i\g\sin(k)}{\sqrt{(\cos(k)-h)^2+\g^2\sin^2(k)}},  
\]
which gives the following generation functions $f_1(x)$, $f_2(x)$ 
for the correlators $\la S^{x}_x S^{x}_0\ra$, $\la S^{y}_x S^{y}_0\ra$ 
respectively for the XY spin chain in the magnetic field: 
\be
f_{1,2}(x)= e^{\mp ix}\sign_{\g}(x)= e^{\mp ix}
\left(\frac{\cos(k)-h+i\g\sin(k)}{\cos(k)-h-i\g\sin(k)}\right)^{1/2}. 
\label{f12}
\ee
Finally, let us obtain the determinant representation for the 
exponential correlator 
\be
G_{\a}(x)=\la e^{i\a N(x)}\ra ,
\label{a}
\ee
where $\a$ is an arbitrary parameter and $N(x)=\sum_{i=1}^{x}n_i$, 
introducing the operators 
\[
A_i=e^{i\a/2}a_i^{+}+a_i, ~~~B_i=a_{i}^{+}+e^{i\a/2}a_i, ~~~
A_i B_i=e^{i\a n_i}, 
\]
for an arbitrary $\a$ we represent the correlator (\ref{a}) as the 
average $\la\prod_{i=1}^{x}(a_i B_i)\ra$. 
Since for $\a\neq\pi$ the averages $\la A_i A_j\ra$ and 
$\la B_i B_j\ra$ are not equal to zero at $\g\neq0$, 
we obtain the correlator 
(\ref{a}) as a Pfaffian, which can be expressed through 
the Toeplitz determinants. 
For the case $\a=\pi$ we have $\la A_i A_j\ra=\la B_i B_j\ra=0$ 
and the average can be represented as the Toeplitz determinant 
$\det_{ij}\left(\la A_i B_j\ra\right)$, $i,j=1, \ldots x$, 
At $\g=0$ the generating function for this determinant equals 
$f(x)=\sign(|x|-p_F)$.

For the practical calculations one should represent the functions 
(\ref{f12}) in the following form: 
\be 
f_{1,2}(x)=\left(e^{i2x}\right)
\left(\frac{1-\l_{1}e^{-ix}}{1-\l_{1}e^{ix}}\right)^{1/2} 
\left(\frac{1-\l_{2}e^{-ix}}{1-\l_{2}e^{ix}}\right)^{1/2}, 
\label{ff12} 
\ee
where the factor $e^{i2x}$ in the parenthesis corresponds to 
the correlator $\la S^{y}_{x}S^{y}_0\ra$ 
(to the function $f_2(x)$) and the parameters $\l_{1,2}$ are 
given by the expression: 
\be
\l_{1,2}=\frac{1}{1+\g}\left(h\mp\sqrt{h^2+\g^2-1}\right). 
\label{l12}
\ee
The behaviour of the correlators is determined by the values of 
the parameters $\l_{1,2}$, which obey the relation 
$\l_1\l_2=a$, where the parameter $a<1$ equals 
$a=(1-\g)/(1+\g)$. 
In general, for $\g>0$ one should distinguish four different cases. 
1) $h^2+\g^2<1$. In this case the parameters 
$\l_{1,2}=\sqrt{a}e^{\pm i\phi}$ are the complex numbers. 
2) $h^2+\g^2>1$ but $h<1$. In this case the parameters 
$\l_{1,2}$ are real and obey $\l_{1}<\l_{2}<1$. 
3) The line $h=1$ at arbitrary $\g$: $\l_{1}=a<1$, $\l_{2}=1$. 
4) The region $h>1$ where the real parameters $\l_{1,2}$ 
are $\l_{1}<1<\l_{2}$. The asymptotic behaviour of the 
correlators is different in this four regions.

\vspace{0.2in}

{\bf 3. Correlators for the XX spin chain.}

\vspace{0.1in}

In this section we calculate the correlators for the isotropic 
XX spin chain ($\g=0$) at non-zero magnetic field. For this purpose 
it is sufficient to use the equation (\ref{det}). 
Let us present the generating functions $f(x)$ for the initial determinants 
of the $x\times x$ matrices obtained for different correlators. 
We construct the functions with the continuously defined argument 
at the interval $(-\pi,\pi)$ which is important for the application of the 
equation (\ref{det}) (see \cite{Basor}). 
One can use the following functions.  
First for the correlator (\ref{a}) at $\a<\pi$ one should use the function 
\be
f(x)=(1,e^{i\a},1)
\label{f1}
\ee
where the three number in parenthesis denote the values of the function 
at the intervals $(-\pi,-p_F)$, $(-p_F,p_F)$ and $(p_F,\pi)$ 
respectively and the unity in eq.(\ref{f1}) denotes the phase equal to zero, 
the function 
\be
f(x)= (1,-1,1)
\label{f2}
\ee
for the same correlator at $\a=\pi$ and the functions of the form 
\be 
f_1(x)=e^{-ix}(e^{-i\pi},1,e^{i\pi}),~~~~ 
f_2(x)=e^{ix}(e^{i\pi},1,e^{-i\pi}), 
\label{f3}
\ee
with the continuously defined argument for the two terms in 
eq.(\ref{G}) respectively. 
Let us note that for the case $\a>\pi$ in eq.(\ref{a}) 
the function can be chosen in the form $(e^{i2\pi},e^{i\a},e^{i2\pi})$, 
which shows that the correlator (\ref{a}) is in fact the 
$2\pi$ - periodic function of the real parameter $\a$.

The function $f(x)$ (\ref{f1}) has the continuously defined argument 
and the two jumps of the size $i\a$ and $-i\a$ at the points 
$x_1=-\pi/2$ and $x_2=\pi/2$.
Substituting this functions into the equation (\ref{det}) we  
obtain at the half-filling the asymptotics: 
\be
G_{\a}(x)=\la e^{i\a N(x)}\ra =e^{i\a x/2}\frac{1}{x^{\a^2/2\pi^2}}
\frac{1}{2^{\a^2/2\pi^2}}\left(g(\l)\right)^2, 
\label{ga}
\ee
where  $\l=\a/2\pi$. 
Evaluating the product in $g(\l)$  for the function given by 
the equation (\ref{gl}) we obtain the expression: 
\be
\ln(g(\l))=\l^2\int_{0}^{\infty}\frac{dt}{t}\left(e^{-t}- 
\frac{1}{\l^2}\frac{(\sh(\l t/2)^2}{(\sh(t/2))^2} 
\right), ~~~~\l^2=\frac{\a^2}{4\pi^2}. 
\label{lgl}
\ee
For the correlator $G_{\a}(x)$ at $\a=\pi$ one should use 
the conjecture (\ref{DN}). 
The function $f(x)=(1,-1,1)$ (\ref{f2}) allows for two different 
representations in the form (\ref{f}) with two equal 
exponents $\sim 1/x^{1/2}$: 1) $b_1=1/2$, $b_2=-1/2$, $f_0(x)=e^{i\pi/2}$ 
and $b_1=-1/2$, $b_2=1/2$, $f_0(x)=e^{-i\pi/2}$, with 
$x_1=-\pi/2$, $x_2=\pi/2$. 
Thus for the asymptotics one should take the sum of the two terms 
which gives the following result: 
\be
G_{\pi}(x)=\left(e^{ip_F x}+e^{-ip_F x}\right)G_0(x)=
\cos(p_{F}x)2G_0(x), 
\label{pi}
\ee
where $p_F=\pi/2$ and 
\[
G_0(x)=\frac{1}{x^{1/2}}\frac{1}{\sqrt{2}}(g(1/2))^2. 
\]
The correlator $G_{\pi}(x)$ equal to zero for $x$ - odd and 
to $\pm 2G_0(x)$ for $x$- even in accordance with the general 
properties of the average $G_{\pi}(x)=\la\prod_{i=1}^{x}(1-2n_i)\ra$ 
(real function).  
 The same result for $G_{\pi}$ can be obtained using the reduction 
of the determinant to the square of the half-size determinant. 
At $\a\neq\pi$ the first term $e^{ip_F x}$ in eq.(\ref{pi}) 
corresponds to the expression (\ref{ga}) while the second term gives 
the subleading asymptotics. 
One can also obtain the asymptotics of the correlator (\ref{pi}) 
at the arbitrary $p_F\neq\pi/2$: 
\be
G_{\pi}(x)=(e^{ip_F x}+e^{-ip_F x})\frac{1}{x^{1/2}}
\frac{1}{\sqrt{\sin(p_F)}}\frac{1}{\sqrt{2}}(g(1/2))^2, 
\label{pf}
\ee
which also have two different exponents and at arbitrary $p_F$ is the 
real function of $x$. 
Note that the same results could be obtained from equation (\ref{det}) 
by taking the sum of the two terms corresponding to the functions 
$(1,e^{i\pi},1)$, and $(1,e^{-i\pi},1)$.

The spin-spin correlator (\ref{corr}) using the 
functions (\ref{f3}) for the two terms we obtain from eq.(\ref{det}) 
the asymptotics 
\be
G(x)=\frac{1}{x^{1/2}}\frac{1}{\sqrt{2}}(g(1/2))^2, 
\label{asg}
\ee
where the constant $g(1/2)=\sqrt{\pi}(G(1/2))^2$, 
can be represented in the form:
\[
\ln(g(1/2))=\frac{1}{4}\int^{\infty}_{0}\frac{dt}{t}\left( 
e^{-4t}-\frac{1}{(\ch(t))^2}\right). 
\]
Note that for the functions (\ref{f3}) there is only one representation 
(\ref{f}), $b_1=b_2=1/2$ and $b_1=b_2=-1/2$ respectively 
so the conjecture (\ref{DN}) leads to the same results. 

Finally, one can use equation (\ref{G}) and the functions (\ref{f3}) 
and the discontinuities at the points $x_{1,2}=\pm p_F$ to obtain 
the asymptotics of the correlator $G(x)$ for the XX spin chain in the 
magnetic field in the form: 
\be
G(x)=\frac{1}{x^{1/2}} \frac{\sqrt{\sin(p_F)}}{\sqrt{2}} 
(g(1/2))^{2}, 
\label{PF}  
\ee
which differs from the well known asymptotics (\ref{asg}) 
by the simple factor $(\sin(p_F))^{1/2}=(1-h^2)^{1/4}$. 
Note that both eq.(\ref{DN}) and (\ref{det}) predict the 
real positive function $G(x)$ without the oscillating factors 
in agreement with the observation that the ground-state 
wave function of the Hamiltonian (\ref{H}) at $\g=0$ 
is the positive function of the coordinates.

\vspace{0.2in}

{\bf 4. Correlators for the XY spin chain.}

\vspace{0.1in}

Let us start with the calculations of the asymptotics of the 
$xx$-correlator $G^{xx}(x)=\la S^{x}_{x}S^{x}_0\ra$ for the 
different values of the parameters $\g$, $h$ ($\l_{1,2}$). 
In Section 2 this correlator was represented by the 
determinant corresponding to the function $f_1(x)$ (\ref{ff12}). 
At $h<1$ we have for the parameters $\l_{1,2}$ the condition 
$|\l_{1,2}|<1$, and this function is the smooth function with 
zero winding number, so the strong Szego theorem can be applied. 
Thus in the whole region $h<1$ the correlator has the long-range 
order which can be easily calculated: 
\[
G^{xx}(x)={1\over4}(1-a^2)^{1/2}(1-h^2)^{1/4}, ~~~~~
a=\frac{1-\g}{1+\g}, ~~~~~~h<1. 
\]
Next, let us consider the correlator at the line $h=1$. 
At this value of $h$ the parameter $\l_2=1$ and the function 
$f_1(x)$ takes the form: 
\[
f_1(z)=
\left(\frac{1-a/z}{1-az}\right)^{1/2} 
\left(\frac{1-1/z}{1-z}\right)^{1/2}, ~~~~~z=e^{ix}. 
\]
One can see that this function has the form (\ref{f}) 
with the function $f_0(x)$ given by the first factor in the last 
equation and the single jump discontinuity at $z=1$ with the 
parameter $b=-1/2$. The calculations according to the formulas 
(\ref{DN}), (\ref{E}) yield the result  
\[
G^{xx}(x)=\frac{1}{x^{1/4}}\frac{\g^{3/4}}{2(1+\g)}
\sqrt{\pi}(G(1/2))^2, ~~~~~~h=1, 
\]
in agreement with the result of ref.\cite{MC}. 
Let us turn to the case $h>1$. In this case $\l_2>1$ and the 
function takes the following form: 
\be
f_1(z)=z^{-1}
\left(\frac{1-\l_1/z}{1-\l_1 z}\right)^{1/2}
\left(\frac{1-\l z}{1-\l/z}\right)^{1/2}, 
~~~~~~\l=1/\l_2<1. 
\label{fun}
\ee
To calculate the determinant we use the method proposed in 
ref.\cite{F}, which allows one to use the general 
Fisher-Hartwig conjecture (\ref{DN}), (\ref{E}). 
Representing the matrix elements as an integrals 
in the complex plane over the unit circle, 
we deform the contour of integration to the circle $|z|=1/\l$ 
which is the {\it outer} boundary of the annulus $\l<|z|<1/\l$. 
This procedure is correct since the function (\ref{fun}) 
is an analytic function inside this annulus and reduces 
to the substitution $z\to z/\l$ in eq.(\ref{fun}). 
As a result we obtain the function of the form (\ref{f}): 
\[
f_0(z)=\l\left(\frac{1-\l_1\l/z}{1-(\l_1/\l)z}\right)^{1/2}
\frac{1}{(1-\l^2/z)^{1/2}},~~~~~a=\frac{1}{4},~~~~b=-\frac{3}{4} 
\]
with the singularity at the point $z=1$.  
Applying the conjecture (\ref{DN}) we obtain the result 
for the case $h>1$: 
\[
G^{xx}(x)=\frac{1}{x^{1/2}}\left(\frac{1}{\l_2}\right)^{x}
\frac{1}{4\sqrt{\pi}}(1-\l_{1}^2)^{1/4} (1-\l_{2}^{-2})^{-1/4} 
(1-\l_1\l_2)^{1/2}, ~~~~~~h>1,  
\]
in agreement with the corresponding formula in ref.\cite{MC}.

Let us proceed with the calculations of the $yy$-correlator 
$G^{yy}(x)=\la S^{y}_{x}S^{y}_0\ra$. 
The behaviour of this correlator in the four regions  
in the space of parameters is completely different.  
We begin with the basic region in the space of parameters 
where both $\g$ and $h$ are sufficiently small: 
$h^2+\g^2<1$. In this case the parameters $\l_{1,2}$ are 
inside the unit circle at the complex plane: 
\[
\l_{1,2}=(1+\g)^{-1}\left(h\pm i\sqrt{1-h^2-\g^2}\right)=  
e^{\pm i\phi}\sqrt{a}. 
\]
In order to obtain an appropriate representation (\ref{f}) 
we deform the contour of integration from the unit circle 
to the circle of the radius $\sqrt{a}$, which amounts to 
the substitution $z\to\sqrt{a}z$ in the function $f_2(z)$ 
(\ref{ff12}). One can see that the resulting function $f(z)$ 
has the two singularities at the points $z=e^{\pm i\phi}$. 
One can prove that the only possible way to represent this 
function in the form (\ref{f}) is as follows. Namely, 
we single out the two equivalent Fisher-Hartwig singularities 
at the points $x=\pm\phi$. Thus we have the representation: 
\[
f_0(z)=a(1-e^{i\phi}az)^{-1/2}(1-e^{-i\phi}az)^{-1/2}, 
~~~~~a_1=a_2=\frac{1}{4}, ~~~~~ b_1=b_2=\frac{3}{4}. 
\]
Substituting this function into the equations (\ref{DN}), 
(\ref{E}) we obtain the correlator: 
\[
G^{yy}(x)=a^{x}\frac{1}{x}\left(\frac{1}{2\pi} 
\sin(\phi)(1-a)^{-1/2}(1+a^2-2a\cos(2\phi))^{-1/4}\right). 
\]
 Next, let us consider the case $h^2+\g^2>1$, but $h<1$, 
when the parameters $\l_1$ and $\l_2$ are real and both are 
inside the unit circle, $\l_1<\l_2<1$. 
Taking the contour to be the inner boundary of the region 
$\l_2<|z|<1/\l_2$, where the function $f_2(z)$ (\ref{ff12}) 
is an analytic function, we obtain the representation: 
\[
f_0(z)=\l_{2}^{2} 
\left(\frac{1-(\l_1/\l_2)/z}{1-(\l_1\l_2)z}\right)^{1/2}
\frac{1}{(1-\l_2^2 z)^{1/2}}, 
~~~~~a=\frac{1}{4},~~~~b=\frac{7}{4},  
\]
which gives the following correlation function: 
\[
G^{yy}(x)=-\frac{1}{x^{3}}\left(\l_2^2\right)^{x}
\frac{1}{8\pi}(1-\l_{1}^2)^{1/4}(1-\l_{2}^{2})^{-3/4} 
(1-\l_1\l_2)^{-1/2}(1-\l_1/\l_2)^{-1},  ~~~~~h<1.   
\]
This expression is identical with that given in ref.\cite{MC}  
which can be easily established using the following 
useful relations: $1-\l_{1}\l_2=1-a=2\g/(1+\g)$ and  
$(1-\l_1^2)(1-\l_2^2)=4(1-h^2)/(1+\g)^{2}$.

As in the case of the $xx$-correlator, at the line $h=1$ 
no special deformation of the integration contour is 
required. The parameters $\l_{1,2}$ are $\l_1=a$, $\l_2=1$ 
and from the function $f_2(x)$ (\ref{ff12}) 
one can easily obtain the representation: 
\[
f_0(z)=\left(\frac{1-a/z}{1-az}\right)^{1/2},  
~~~~~a=0,~~~~b=\frac{3}{2},  
\]
Calculating the constant (\ref{E}) we get: 
\[
G^{yy}(x)=-\frac{1}{x^{9/4}}\frac{\sqrt{\pi}}{32} 
\g^{-5/4}(1+\g)(G(1/2))^2.  
\]
Finally, let us consider the region $h>1$ ($\l_2>1$). 
In this case the calculations are analogous to the 
corresponding case for the $xx$-correlator. 
Introducing the parameter $\l=1/\l_2<1$, we  
observe that the function is an analytic function 
in the annulus $\l<|z|<1/\l$. Deformation of the integration 
contour to the {\it inner} boundary of the annulus, 
$z\to\l z$, gives the representation (\ref{f}) 
in the form: 
\[
f_0(z)=\l\left(\frac{1-(\l_1/\l)/z}{1-\l_1\l z}\right)^{1/2}
(1-\l^{2}z)^{1/2},~~~~~a=-\frac{1}{4},~~~~b=\frac{5}{4}. 
\]  
Applying the conjecture (\ref{DN}) we obtain the result:  
\[
G^{yy}(x)=-\frac{1}{x^{3/2}}\left(\frac{1}{\l_2}\right)^{x}
\frac{1}{8\sqrt{\pi}}(1-\l_{1}^2)^{1/4}(1-\l_{2}^{-2})^{3/4} 
(1-(\l_1/\l_2))^{-1}(1-\l_1\l_2)^{-1/2}, 
~~~~h>1,  
\]
in agreement with the corresponding formula in ref.\cite{MC}. 
Let us note that in all the cases considered in this section 
(except the $yy$- correlator at $h^2+\g^2<1$) we have 
obtained the functions with the single Fisher-Hartwig 
singularity. In the case of the single singularity 
the asymptotics (\ref{DN}), (\ref{E}) was proved 
in ref.\cite{ES}. 
In conclusion, it is easy to verify that in all the cases 
considered, there is only one possible way to choose the contour 
of integration (inner or outer boundary of the annulus). 
In fact, in all the cases at the opposite boundary the 
condition $a_r\pm b_r\in Z^{-}$ is fulfilled. 
Formally, in this case we have the constant $E=0$, 
which suggests, that these special cases should be excluded from 
the general statement of the conjecture (\ref{DN}), (\ref{E}) 
since in this case the matrix elements corresponding to the 
pure Fisher-Hartwig singularity does not exist.

\end{document}